 \documentclass[final,5p,times,twocolumn]{elsarticle}

\usepackage{amsmath,amssymb}
\usepackage{amsthm}
\usepackage{graphicx}
\usepackage{xurl}
\usepackage[colorlinks=true,breaklinks=true]{hyperref}
\usepackage{diagbox}
\usepackage{makecell}

\journal{Physics Letters B}

\begin{document}

\begin{frontmatter}






\title{Revisiting neutron-skin thickness and dipole polarizability constraints on the symmetry energy in Antisymmetrized Molecular Dynamics}





\author[label1]{Dandan Niu}
\author[label1]{Luqi Li}
\author[label1]{Xinyu Wang}
\author[label1]{Ping Feng}
\author[label1]{Qiang Zhao}
\author[label1]{Kai Zhao}
\author[label2]{Akira Ono}
\author[label1,label3,label4]{Yingxun Zhang\corref{cor1}}
\ead{zhyx@ciae.ac.cn}
\affiliation[label1]{China Institute of Atomic Energy, P. O. Box 275(18), Beijing 102413, China}
\affiliation[label2]{Department of Physics, Graduate School of Science, Tohoku University, Sendai 980-8578, Japan}
\affiliation[label3]{Department of Physics and Technology, Guangxi normal University, Guilin 540101, China}
\affiliation[label4]{Southern Center for Nuclear-Science Theory (SCNT), Institute of Modern Physics, Chinese Academy of Sciences, Huizhou 516000, China}

\cortext[cor1]{Corresponding author}

\begin{abstract}

The neutron-skin thickness and electric dipole polarizability are among the most sensitive probes of the symmetry energy at subsaturation densities. Motivated by the tension raised by recent analyses of PREX-II and CREX data within density-functional-based approaches, we perform a unified study of static and dynamical isovector observables within the antisymmetrized molecular dynamics (AMD) framework. Using thirty interaction parameter sets that span different values of the symmetry-energy coefficient $S_0$, slope parameter $L$, and neutron-proton effective-mass splitting $\Delta m_{np}^*$, we systematically analyze the neutron-skin thicknesses of nuclei from $^{40}$Ca to $^{238}$U together with the electric dipole polarizability $\alpha_D$ of $^{208}$Pb. A combined $\chi^2$ analysis of neutron-skin thicknesses and the electric dipole polarizability yields preferred values of $L$ that increase with $S_0$, reflecting the joint constraint from the static and dynamical observables. 
Furthermore, 
we identify the density region mainly probed by these observables as $0.019\le \rho/\rho_0\le 0.60$, where the relative narrowing strength function varies by less than 10\% compared to its maximum narrowing strength. The maximum reduction of the uncertainty of $S(\rho)$ occurs at $0.28\rho_0$, where the symmetry energy within 1$\sigma_{\rm post}$ uncertainty is constrained to be $S(0.28\rho_0)=13.84\pm 1.31$ MeV.  
These results demonstrate that a unified AMD analysis of neutron-skin systematics and dipole polarizability provides a complementary constraint on the symmetry energy below saturation density.

\end{abstract}


\end{frontmatter}


\section{Introduction}

The nuclear symmetry energy characterizes the isospin dependent part of the energy for isospin asymmetric nuclei, $a_{sym}(N-Z)^2/A$, or for isospin asymmetric nuclear matter, $S(\rho)\delta^2$~\cite{BALi08PR}. It plays a crucial role in our understanding of the properties of neutron stars~\cite{BALi21Universe,Burgio21PPNP,Tsang24nature,Lattimer14EPJ,NBZhang18APJ}, heavy-ion collision mechanisms and observables~\cite{BALi08PR,YXZhang14PLB,LWChen05PRL}, and the ground and collective excited state properties of nuclei~\cite{ROCA18PPNP,Tsang12PRC,Brown2000PRL,TGYue22PRR,LGCao15PRC,Centelles09PRL,Reinhard21PRL,WBHe14PRL,Roca11PRL,ZZhang13PLB}. However, theoretical predictions for the density dependence of the symmetry energy $S(\rho)$ still have large uncertainties away from the normal density. Therefore, constraining $S(\rho)$ has become one of the main goals of nuclear physics~\cite{NUPECC24LRP,USDOE23LRP}.

Considerable efforts have been devoted to constraining the nuclear equation of state and the nuclear symmetry energy at different densities via different isospin sensitive observables~\cite{Colo14EPJ,Reinhard10PRC,Roca13PRC,HZheng16PRC,Klimkiewicz07PRC,Carbone10PRC,Vretenar12PRC}, and comprehensive reviews of this subject can be found in Refs.~\cite{BALi21Universe,Tsang12PRC}. Recent high-precision measurements, such as those from the Lead Radius Experiment II (PREX-II)~\cite{Adhikari21PRL} and the Calcium Radius Experiment (CREX)~\cite{Adhikari22PRL}, have provided new results on neutron-skin thicknesses, and analyses based on different density functional theory (DFT) frameworks have revealed tensions in the extracted symmetry energy parameters~\cite{Reed21PRL}. 
Very recently, in studies based on extended nonrelativistic Skyrme energy density functionals (EDFs) and covariant energy density functionals, a strong isovector spin-orbit interaction has been proposed as a promising solution~\cite{TGYue26Science,kunji}. However, such a strong isovector spin-orbit interaction destroys the agreement with the established spin-orbit phenomenology, primarily by modifying the well-known ordering of spin-orbit partners~\cite{kunji}. This situation highlights the need for independent theoretical approaches and systematic analyses beyond the energy density functional framework, which can help assess model dependencies and provide complementary constraints.

Antisymmetrized molecular dynamics (AMD)~\cite{Ono92PRL,Ono92PTP} provides such an alternative by offering a fully microscopic description of nuclear many-body systems without relying on predefined mean-field potentials. In AMD, nucleons are described as localized wave packets, and many-body correlations, including cluster degrees of freedom and dynamical fluctuations, can be incorporated naturally. This feature is particularly important for neutron-rich nuclei, where surface properties and isovector dynamics are closely related to neutron skin formation and dipole response, both of which can be sensitive to correlations beyond the mean-field approximation. Furthermore, the time-dependent formulation of AMD enables a unified and self-consistent description of ground-state properties and dynamical responses, such as electric dipole excitations, within the same framework~\cite{DDNiu26arXiv}. 

In this work, we employ the AMD model~\cite{Ono92PRL,Ono92PTP} to systematically analyze neutron-skin thicknesses from $^{40}$Ca to $^{238}$U and the electric dipole polarizability of $^{208}\rm Pb$. The purpose of using AMD is not merely to provide another numerical calculation of $\Delta R_{np}$ and $\alpha_D$. Its central role is to test whether the static neutron and proton density distributions and the dynamical isovector restoring force can be described within the same time-dependent microscopic framework and by the same density dependence of the symmetry energy. This provides a test complementary to energy density functional (EDF) analyses combined with the random-phase approximation (RPA), in which the ground state and the response are usually connected through a small-amplitude linearization around a mean-field solution.

The paper is organized as follows. First, we briefly describe the effective interaction used to calculate neutron-skin thicknesses and electric dipole polarizability. The main results and discussion are presented in the following section, which includes the calculated neutron-skin thicknesses of nuclei from $^{40}$Ca to $^{238}$U, the electric dipole polarizability of $^{208}\rm Pb$, the sensitive density region probed by these observables, and the resulting constraint on the symmetry energy. Finally, we give a summary and outlook.

\section{Model}\label{Sec:Theoretical framework}


The AMD model used in this work is the same as in Refs.~\cite{Natsumi16PRC,Natsumi23PRC}. 
In the present calculations, we adopt the Skyrme-type effective interaction with the spin-orbit term omitted:
\begin{equation}
\begin{aligned}
    v_{ij}=&t_0(1+x_0P_{\sigma})\delta(\mathbf{r})+\frac{1}{2}t_1(1+x_1P_{\sigma})[\delta(\mathbf{r})\mathbf{k}'^2+\mathbf{k}^2\delta(\mathbf{r})]\\
    &+t_2(1+x_2P_{\sigma})\mathbf{k}'\cdot \delta(\mathbf{r})\mathbf{k}+\frac{1}{6}t_3(1+x_3P_{\sigma})[\rho(\mathbf{r}_i)]^{\gamma}\delta(\mathbf{r})\\
    &+v_\rho^{ext}
\end{aligned}
\end{equation}
where $\mathbf{r}=\mathbf{r}_i-\mathbf{r}_j$ and $\mathbf{k}=\frac{1}{2\hbar}(\mathbf{p}_i-\mathbf{p}_j)$. The additional term $v_\rho^{ext}=\frac{1}{6}t_3x_3'P_\sigma\delta(\mathbf{r}_1 - \mathbf{r}_2)(\rho(\mathbf{r}_1)^\gamma -\rho_0^\gamma)$ was introduced to vary $S_0$ and $L$, while keeping the incompressibility $K_0$, the isoscalar effective mass $m_s^*$, and the isovector effective mass $m_v^*$ fixed~\cite{Natsumi16PRC}.



With the additional density-dependent interaction term, the expression for the symmetry energy $S(\rho)$ becomes

\begin{equation}
    \begin{aligned}
        S(\rho)&=\frac{1}{3}\frac{\hbar^2}{2m}\bigg(\frac{3\pi^2}{2}\bigg)^{2/3}\rho^{2/3}-\frac{1}{8}t_0(2x_0+1)\rho\\
        &\quad-\frac{1}{24}\bigg(\frac{3\pi^2}{2}\bigg)^{2/3}(3\Theta_v-2\Theta_s)\rho^{5/3}- \frac{1}{48}t_3( 2x_3 \rho_0^{\gamma}\rho+\rho^{\gamma+1} )\\
        &\quad- \frac{1}{24}t_3 x_3' \left( \rho^{\gamma+1} - \rho_0^{\gamma}\rho \right)
    \end{aligned}
\end{equation}
where $\Theta_s=3t_1+t_2(5+4x_2)$ and $\Theta_v=t_1(2+x_1)+t_2(2+x_2)$.

\begin{figure}[htbp]
	\centering
	\includegraphics[width=0.8\linewidth]{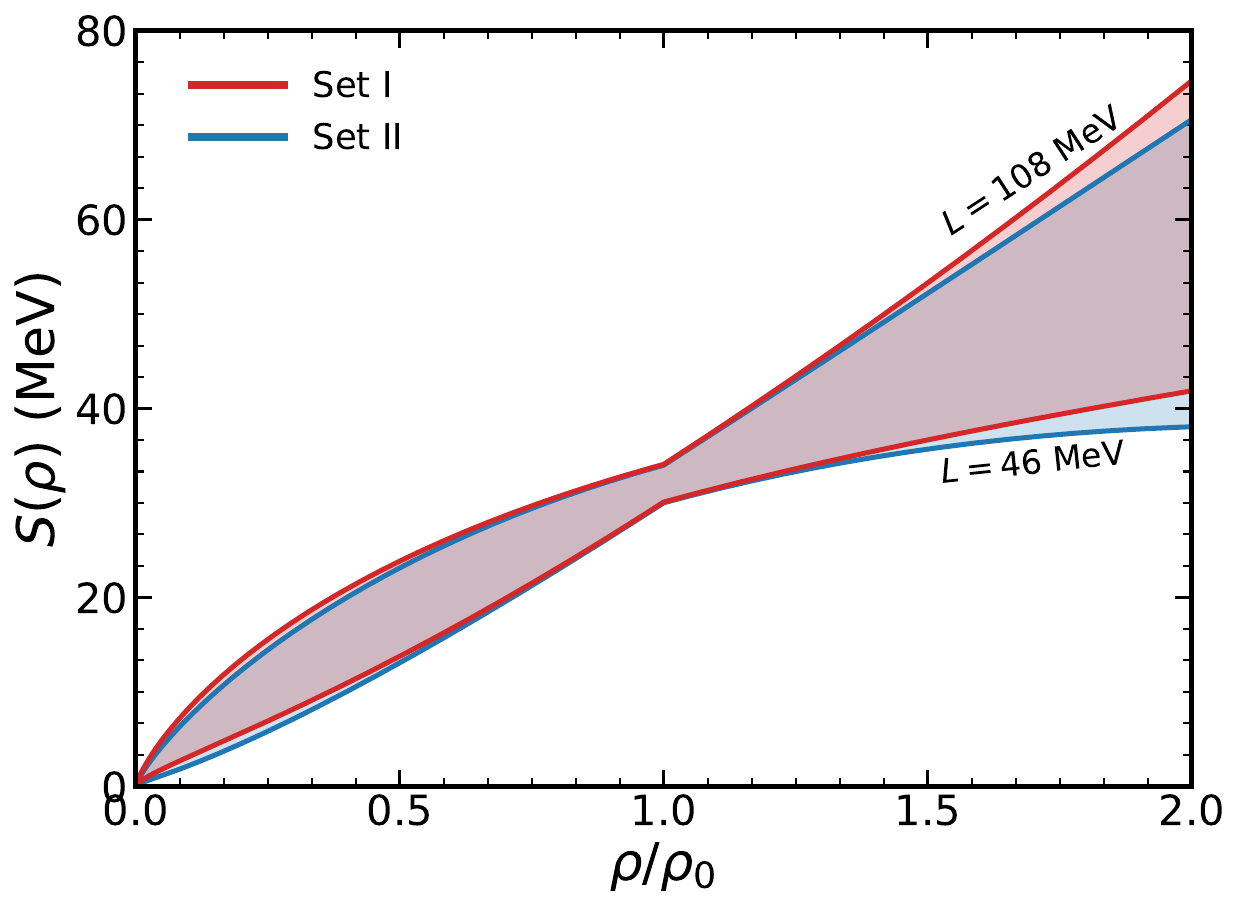}
	\caption{Density dependence of the symmetry energy $S(\rho)$ for the 30 parameter sets used in this work.}
    \label{fig:symmetry energy}
\end{figure}

In the following calculations, we vary $S_0$ and $L$ at given $m_s^*$ and $m_v^*$. In total, 30 parameter sets are used. The corresponding values are presented in Table~\ref{tab:parameter_sets}, and the density dependence of the symmetry energy $S(\rho)$ is displayed in Fig.~\ref{fig:symmetry energy}. The shaded regions indicate the ranges of $S(\rho)$ spanned by the parameter sets. The red and blue bands correspond to Set I and Set II, respectively. Set I corresponds to parameter sets with $m_n^*<m_p^*$, and Set II corresponds to those $m_n^*>m_p^*$. By comparing the calculated results from these sets, one may understand not only the influence of $S_0$ and $L$ but also the effect of the effective mass splitting $\Delta m_{np}^*$. 



\begin{table}[htbp]
  \caption{Parameter sets used in the calculations, $S_0$ and $L$ are in MeV.}
  \label{tab:parameter_sets}
  \setlength{\tabcolsep}{3pt}
  \begin{tabular}{l c c c c c } 
  \hline
  \hline
  Sets&$m_s^*/m$ & $m_v^*/m$ & $S_0$& $L$ & $\Delta m_{np}^*$\\
  \hline
  I&$0.695$ & $0.800$ & [30, 32, 34] & [46, 61, 75, 92, 108]  & $<0$\\
  II&$0.789$ & $0.653$ & [30, 32, 34] & [46, 61, 75, 92, 108]  & $>0$\\
  \hline
  \hline
  \end{tabular}
\end{table}


To find the ground state of the nucleus, the frictional cooling method is adopted~\cite{Ono92PRL,Ono92PTP,DDNiu26arXiv} by finding its minimum ground-state energy at the given effective interaction. 
However, one should note that there are two extreme cases in AMD for the true ground state. One is that the true ground-state wave function contains many configurations, it is reasonable to average the final observables over different configurations, as in the typical case of a well-deformed nucleus. On the other hand, if the cooling is trapped in local minima that do not correspond to the ground state, those configurations should not, in principle, be included, and therefore the configuration dependence should be regarded as indicating uncertainties originating from the model. The actual case may be a mixture of these extremes. For a conservative estimate, fluctuations due to the ground-state configurations are included in the uncertainties of the results we studied observables, i.e., $\Delta R_{np}$, and $\alpha_D$, leading to a more conservative constraint on the symmetry energy.

\section{Results}

In the following, we analyze two classes of isovector observables within the same AMD framework: neutron-skin thicknesses $\Delta R_{np}=\langle r_n^2\rangle^{1/2}-\langle r_p^2\rangle^{1/2}$ for 16 nuclei from $^{40}\mathrm{Ca}$ to $^{238}\mathrm{U}$\footnote{The nuclei considered are $^{40}\mathrm{Ca}$, $^{48}\mathrm{Ca}$, $^{54}\mathrm{Fe}$, $^{56}\mathrm{Fe}$, $^{57}\mathrm{Fe}$, $^{58}\mathrm{Ni}$, $^{59}\mathrm{Co}$, $^{60}\mathrm{Ni}$, $^{64}\mathrm{Ni}$, $^{112}\mathrm{Sn}$, $^{116}\mathrm{Sn}$, $^{124}\mathrm{Sn}$, $^{128}\mathrm{Te}$, $^{208}\mathrm{Pb}$, $^{209}\mathrm{Bi}$, and $^{238}\mathrm{U}$.}, and the electric dipole polarizability $\alpha_D$ of $^{208}\mathrm{Pb}$. Their combined analysis provides a unified constraint on the symmetry energy in the subsaturation-density region.

\begin{figure}[htbp]
	\centering
	\includegraphics[width=1.0\linewidth]{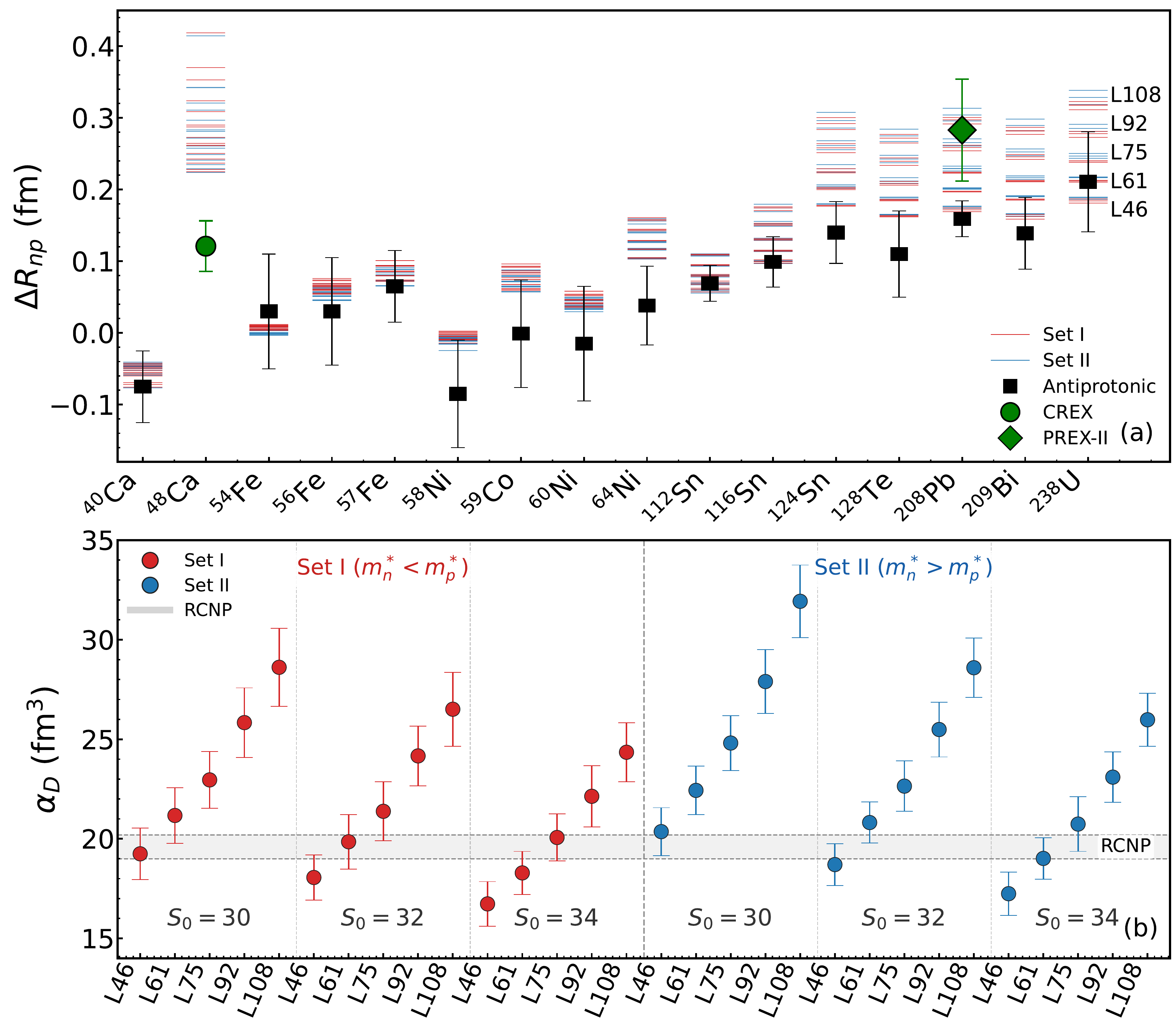}
	\caption{(a) Calculated neutron-skin thicknesses $\Delta R_{np}$ for the nuclei considered in this work. Experimental data from antiprotonic atoms~\cite{Trzciska01PRL,Centelles09PRL}, CREX~\cite{Adhikari22PRL}, and PREX-II~\cite{Adhikari21PRL} are shown for comparison. (b) Calculated electric dipole polarizability $\alpha_D$ of $^{208}\rm Pb$ as a function of $L$ for $S_0=30,32,34\rm \,MeV$ and different sign of $\Delta m_{np}^*$.}
    \label{fig:Rnp-alphaD}
\end{figure}

Fig.~\ref{fig:Rnp-alphaD}(a) shows the calculated neutron-skin thicknesses $\Delta R_{np}$, which are obtained by averaging over 100 configurations, for nuclei from $^{40}\mathrm{Ca}$ to $^{238}\mathrm{U}$. The variance is estimated by $\sigma_{\Delta R_{np,{\rm th}}}^2=\frac{1}{100}\sum_{k=1}^{100}\left(\Delta R_{np,k}^{\rm th}-\Delta \bar{R}_{np}^{\rm th}\right)^2$. In general, $\Delta R_{np}$ increases with increasing mass number, and this increase becomes more pronounced in medium-heavy and heavy nuclei such as the Sn isotopes, $^{208}\mathrm{Pb}$, $^{209}\mathrm{Bi}$, and $^{238}\mathrm{U}$. For a given effective-mass-splitting group ($m_n^*<m_p^*$ or $m_n^*>m_p^*$), $\Delta R_{np}$ also tends to increase with increasing $L$. This behavior originates from the density dependence of the symmetry energy: a larger $L$ corresponds to a stronger neutron pressure in neutron-rich matter, pushing excess neutrons toward the surface and resulting in a thicker neutron skin. Quantitatively, for nuclei lighter than $^{60}\mathrm{Ni}$, the sensitivity of $\Delta R_{np}$ to $L$ is weak. For nuclei heavier than $^{64}\mathrm{Ni}$, the sensitivity of $\Delta R_{np}$ to $L$ becomes stronger, 
indicating that the $\Delta R_{np}$ in heavy nuclei can provide useful constraints on the symmetry energy. 
A further systematic feature is that, for the same $L$, the parameter sets with $m_n^*>m_p^*$ tend to give a thicker neutron skins than those with $m_n^*<m_p^*$. This indicates that $\Delta R_{np}$ is influenced not only by the bulk density dependence of the symmetry energy characterized by $L$, but also by the momentum-dependent isovector mean field associated with the effective-mass splitting. Therefore, the constraints on the symmetry energy should not simply depend on $S_0$ and $L$, but also on $\Delta m_{np}^*$. 


A comparison with experimental data shows that the present AMD framework captures the overall neutron-skin systematics, but not without tension. In particular, for $^{48}\rm Ca$ and $^{64}\rm Ni$, none of the present results falls within the experimental uncertainty band. 
These deviations indicate that the present AMD framework and/or parameters still face challenges in simultaneously reproducing neutron-skin thickness data across all nuclei. 
For constraining the symmetry energy, the $\Delta R_{np}$ in the selected nuclei should be strongly correlated with $L$ and less sensitive to the omitted spin-orbit interaction. The theoretical uncertainties associated with the omitted spin-orbit interaction are estimated using Skyrme-Hartree-Fock-Bogoliubov (SHFB)~\cite{Bennaceur05CPC}, and our calculations show that the changes in $\Delta R_{np}$ are less than 3\% for nuclei above $^{116}\rm Sn$. 

Fig.~\ref{fig:Rnp-alphaD}(b) displays the electric dipole polarizability $\alpha_D$ of $^{208}\rm Pb$, which is calculated as
\begin{equation}\label{eq:alphaD}
    \alpha_D=\frac{8\pi}{9}\int_{\omega_{min}}^{\omega_{max}} \frac{S(\omega)}{\omega}d \omega.
\end{equation}
In our calculations, $\omega_{min}=5$ MeV, the same as the experimental lower limit, and $\omega_{max}=60$ MeV is adopted to ensure convergence of the integral. Three systematic trends can be seen directly from the figure. First, for each fixed $S_0$ and $\Delta m_{np}^*$, $\alpha_D$ increases monotonically with increasing $L$. Second, for each fixed $L$ and $\Delta m_{np}^*$, $\alpha_D$ decreases as $S_0$ increases from 30 to 34 MeV. Third, for the same $S_0$ and $L$, Set II ($m_n^*>m_p^*$) generally gives larger $\alpha_D$ values than Set I ($m_n^*<m_p^*$). This feature can be understood in a simple dynamical picture within the transport model, where the main frequency of the electric dipole oscillation depends on the tensor constant, such as $-e\frac{NZ}{Am} \nabla\bigg((\frac{1}{m_s^*}-\frac{1}{m_v^*})\hbar^2k_F^2\bigg)\frac{1}{2}\frac{\nabla\rho}{\rho}$ and the scalar constant, i.e., $e\frac{NZ}{Am} \bigg[4S(\rho)- \frac{2}{3}L(\rho) \bigg] \frac{(\nabla\rho)^2}{\rho^2}$ as in Ref.~\cite{DDNiu26arXiv}. The $\nabla \rho$ tends to zero near the center of nucleus, the contribution is mainly localized at the nuclear surface, where the density and mean-field gradients are large. Thus, the dipole frequency is especially sensitive to the subsaturation behavior of $S(\rho)$ and $L(\rho)$. It means that a larger $L$ leads to a lower characteristic dipole frequency, and thus generally enhances $\alpha_D$.

To constrain the symmetry energy by comparing the calculations with the experimental data, we construct a global $\chi^2$ function from the neutron-skin thicknesses of four nuclei, $^{124}\mathrm{Sn}$, $^{128}\mathrm{Te}$, $^{208}\mathrm{Pb}$ (in this analysis, the PREX-II data are used for $^{208}\mathrm{Pb}$) and $^{238}\mathrm{U}$. These nuclei are selected because their neutron-skin thicknesses are strongly correlated with $L$ and weakly affected by the spin-orbit interaction. 
Another isospin-sensitive observable is the electric dipole polarizability of $^{208}\mathrm{Pb}$. 
The total $\chi^2$ is written as
\begin{equation}\label{eq:chi2_total}
    \chi^2_{\rm total}=\chi^2_{\Delta R_{np}}+\chi^2_{\alpha_D},
\end{equation}
with
\begin{equation}
    \chi^2_{\Delta R_{np}}=\frac{1}{4}\sum_{i=1}^{4}\frac{\left[\Delta R_{np}^{\rm th}(i)-\Delta R_{np}^{\rm exp}(i)\right]^2}{\sigma^2_{\Delta R_{np},{\rm th}}(i)+\sigma^2_{\Delta R_{np},{\rm exp}}(i)},
\end{equation}
and
\begin{equation}
    \chi^2_{\alpha_D}=\frac{\left[\alpha_D^{\rm th}(^{208}{\rm Pb})-\alpha_D^{\rm exp}(^{208}{\rm Pb})\right]^2}{\sigma^2_{\alpha_D,{\rm th}}+\sigma^2_{\alpha_D,{\rm exp}}}.
\end{equation}
The theoretical uncertainty $\sigma_{\alpha_D,{\rm th}}$ is estimated as the standard deviation associated with the configuration dependence of the AMD calculations, in the same way as for $\Delta R_{np}$, while $\sigma_{\Delta R_{np},{\rm exp}}(i)$ and $\sigma_{\alpha_D,{\rm exp}}$ denote the corresponding experimental uncertainties.
This normalization treats neutron-skin systematics and electric dipole polarizability as two classes of observables with comparable statistical weight, avoiding an artificial dominance of the fit by the number of neutron-skin data points.

\begin{figure}[htbp]
	\centering
	\includegraphics[width=1.0\linewidth]{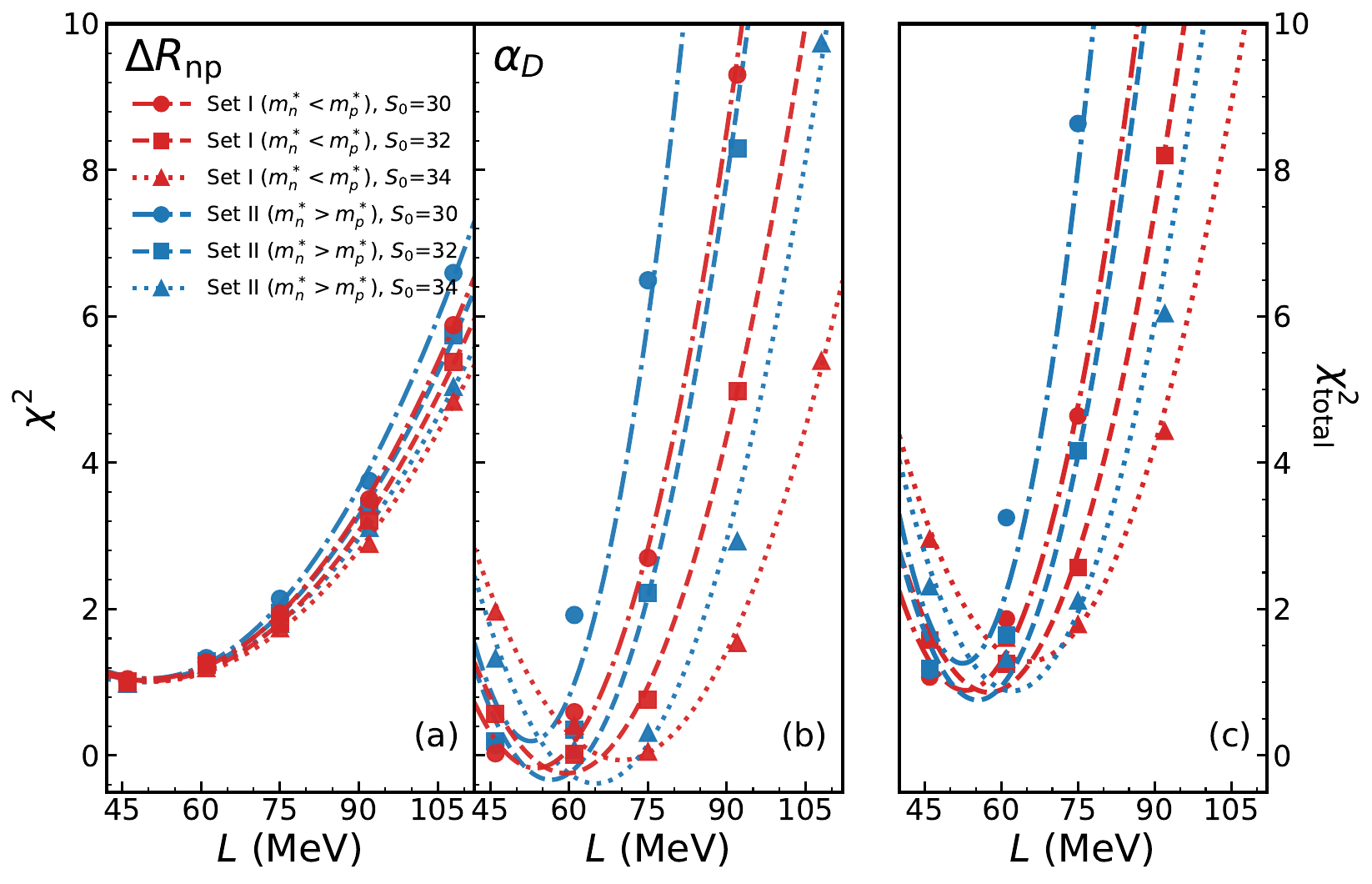}
	\caption{(a) $\chi^2_{\Delta R_{np}}$ and (b) $\chi^2_{\alpha_D}$ as functions of the slope parameter $L$ for different symmetry-energy coefficients $S_0$ and effective-mass-splitting $\Delta m_{np}^*$. Panel (c) shows the total $\chi^2_{\rm total}$ obtained from the combined analysis of the neutron-skin thicknesses and electric dipole polarizability.}
    \label{fig:chi2-analysis}
\end{figure}

Fig.~\ref{fig:chi2-analysis} 
(a) and (b) show the $\chi^2$ as a function of $L$ obtained only from the neutron-skin thickness $\Delta R_{np}$ for nuclei, and the electric dipole polarizability $\alpha_D$ for $^{208}$Pb, respectively. Panel (c) presents the total $\chi^2_{\rm total}$ obtained from the combined analysis as in Eq.(\ref{eq:chi2_total}). 
For each fixed value of $S_0$ and effective mass splitting, the calculated $\chi^2(L)$ values were fitted around the minimum to infer the best-fit value of $L$. 
As shown in Fig.~\ref{fig:chi2-analysis}(a), the systematic neutron-skin analysis shows that the minimum of $\chi^2_{min}\approx 1.0$, corresponds to the slope parameter of the symmetry energy in $L=48.2-50.3~{\rm MeV}$.
The exact values of the constraints are also listed in Table~\ref{tab:L_constraints}. 
Fig.~\ref{fig:chi2-analysis}(b) shows that the $\chi^2$ analysis for the electric dipole polarizability and the $\chi_{min}^2$ depend on the values of $S_0$ and $\Delta m_{np}^*$. Our calculations show that the best-fit values of $L$ are $52.7-69.9~{\rm MeV}$ for the adopted values of $S_0$.

\begin{table}[htbp]
  \caption{
  Constraints on the slope parameter $L$ for different fixed values of $S_0$ and neutron--proton effective-mass splitting. All values are in MeV.
  }
  \label{tab:L_constraints}
  \setlength{\tabcolsep}{3pt}
  {
  \begin{tabular}{l c c c}
  \hline
  \hline
\diagbox{Set}{Obs.}   & $\Delta R_{np}$($\chi_{min}^2$) & $\alpha_D$ ($\chi_{min}^2$) & \begin{tabular}{c}
     $\Delta R_{\rm np} +\alpha_D$ \\
     ($\chi_{min}^2$)
   \end{tabular} \\
  \hline
  $(m_n^*<m_p^*$,$S_0=30)$ &  \makecell[c]{$50.1\pm26.4$\\$(1.04)$} & \makecell[c]{$53.7\pm12.3$\\$(0.00)$} & \makecell[c]{$53.0\pm11.1$\\$(0.89)$}\\
  $(m_n^*<m_p^*$,$S_0=32)$ &  \makecell[c]{$50.3\pm27.7$\\$(1.02)$} & \makecell[c]{$59.5\pm14.2$\\$(0.00)$} &  \makecell[c]{$57.6\pm12.6$\\$(0.86)$} \\
  $(m_n^*<m_p^*$,$S_0=34)$ &  \makecell[c]{$50.2\pm29.7$\\$(1.01)$} & \makecell[c]{$69.9\pm16.4$\\$(0.00)$} &  \makecell[c]{$65.3\pm14.4$\\$(1.28)$}\\
  $(m_n^*>m_p^*$,$S_0=30)$ &  \makecell[c]{$49.9\pm24.8$ \\ $(1.05)$} & \makecell[c]{$52.7\pm9.3$\\$(0.20)$} & \makecell[c]{$52.4\pm8.7$\\$(1.26)$}\\
  $(m_n^*>m_p^*$,$S_0=32)$ &  \makecell[c]{$49.3\pm27.2$\\$(1.03)$} & \makecell[c]{$56.5\pm11.7$\\$(0.00)$} &  \makecell[c]{$55.4\pm10.7$\\$(0.76)$}\\
  $(m_n^*>m_p^*$,$S_0=34)$ &  \makecell[c]{$48.2\pm29.8$\\$(1.00)$} & \makecell[c]{$65.0\pm13.7$\\$(0.00)$} &  \makecell[c]{$62.1\pm12.4$\\$(0.88)$}\\
  \hline
  \hline
  \end{tabular}
  }
\end{table}



Fig.~\ref{fig:chi2-analysis}(c) shows $\chi^2_{\rm total}$ as a function of $L$ for given $S_0$ and $\Delta m_{np}^*$. 
The corresponding constraints on $L$ are summarized in Table~\ref{tab:L_constraints}. For $\Delta m_{np}^*<0$, the minima of $\chi^2_{\rm total}$ are obtained at $L=53.0~{\rm MeV}$, $57.6~{\rm MeV}$, and $65.3~{\rm MeV}$ for $S_0=30$, 32, and 34 MeV, respectively. For $\Delta m_{np}^*>0$, the corresponding best-fit values are $L=52.4~{\rm MeV}$, $55.4~{\rm MeV}$, and $62.1~{\rm MeV}$ for $S_0=30$, 32, and 34 MeV, respectively. 
The monotonic shift of $L_{\rm best}$ toward larger values with increasing $S_0$ indicates a positive correlation between these two symmetry-energy parameters in reproducing the full set of observables simultaneously. 
In other words, a larger symmetry energy at saturation density tends to require a larger slope parameter in order to maintain an overall consistent description of both static and dynamical isovector observables. 

\begin{figure}[htbp]
	\centering
	\includegraphics[width=0.8\linewidth]{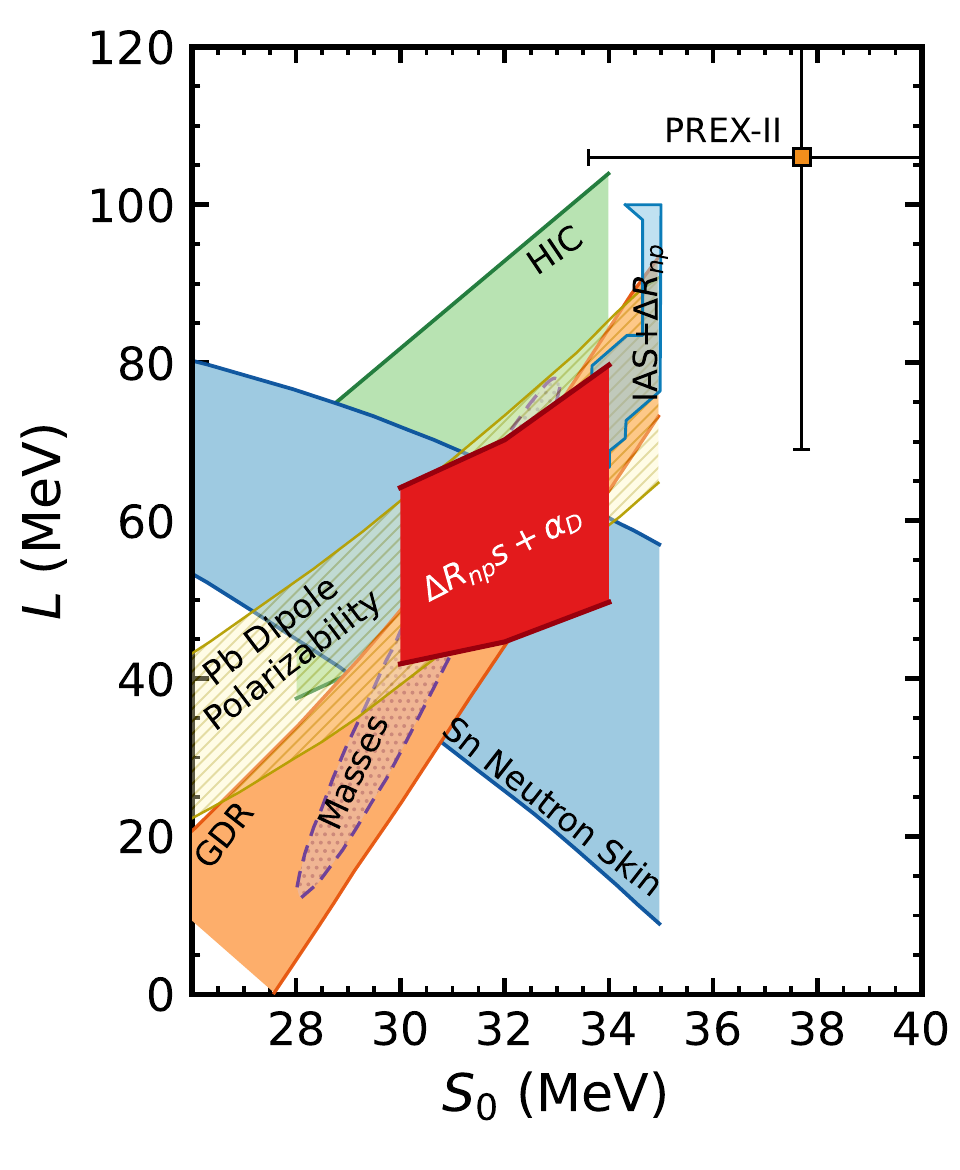}
	\caption{Constraints on the $S_0-L$ correlation obtained in this work (red region), and from a variety of experimental and theoretical approaches. The figure was modified from Refs.~\cite{Lattimer12AR,Reed21PRL}. 
}
    \label{fig:S0-L}
\end{figure}

Fig.~\ref{fig:S0-L} presents this constraint region in the $S_0-L$ plane, which is 
constructed by connecting the $1\sigma$ intervals of $L$ across the explored values of $S_0$. Within the present AMD analysis, the preferred values of $S_0$ and $L$ show a positive correlation: a larger symmetry energy at saturation density is favored together with a larger slope parameter in order to reproduce the combined set of observables.  
It is also useful to compare the present red region with other commonly discussed constraints in the $S_0-L$ plane, as summarized in Fig.~\ref{fig:S0-L}. The displayed regions include constraints derived from giant dipole resonances (GDR)~\cite{Trippa08PRC}, nuclear masses~\cite{Kortelainen10PRC}, dipole polarizability~\cite{Tamii11PRL,Roca13PRC}, neutron-skin data of Sn isotopes~\cite{LWChen10PRC}, heavy-ion collision observables (HIC)~\cite{Tsang09PRL}, and the combined analysis of isobaric analog states with neutron-skin information ($\rm IAS+\Delta R_{np}$)~\cite{Pawel17NPA}. These different observables populate different regions in the $S_0-L$ plane, reflecting their different density sensitivities and underlying model assumptions. Within the explored $S_0$ interval, the present result favors a moderate range of $L$ and provides an AMD-based constraint on the symmetry-energy parameters that can be directly compared with these existing determinations. 
For reference, the PREX-II constraint is also shown~\cite{Adhikari21PRL}. It lies outside most other constraints and highlights the importance of independent neutron-skin measurements, such as those anticipated from the proposed Mainz Radius Experiment (MREX) at the Mainz Energy-Recovering Superconducting Accelerator (MESA)~\cite{Piekarewicz26MREX}.

An important question is which density region is most effectively constrained by the combined analysis. In this work, we determine the sensitive density region 
by a relative narrowing strength $W(\rho)$, defined as
\begin{equation}
    W(\rho)=\frac{\sigma_{\rm prior}(\rho)-\sigma_{\rm post}(\rho)}{\sigma_{\rm prior}(\rho)}.
\end{equation}
This quantity reflects the ability of the isospin-sensitive observables to reduce the uncertainty of $S(\rho)$ relative to the prior uncertainty at a given density. 
Here, $\sigma_{\rm prior}(\rho)$ and $\sigma_{\rm post}(\rho)$ 
are calculated as described in footnote
\footnote{
For each parameter set $k$, a posterior-like weight is assigned as
\begin{equation}\label{eq:weight_wk}
    w_k=\exp\left[-\frac{\chi^2_k-\chi^2_{\min}}{2}\right],
\end{equation}
where $\chi^2_k$ denotes the global $\chi^2_{\rm total}$ value of the set $k$ and $\chi^2_{\min}$ is the minimum value among the sets we used. Within the present discrete ensemble of AMD interaction parameter sets, the weights in Eq.~(\ref{eq:weight_wk}) may be interpreted as relative likelihood weights. 
At each density, the posterior-weighted mean value and standard deviation of the symmetry energy are calculated as
\begin{equation}
    \bar S_{\rm post}(\rho)=\frac{\sum_k w_k S_k(\rho)}{\sum_k w_k},
\end{equation}
and
\begin{equation}
    \sigma_{\rm post}(\rho)=\left[\frac{\sum_k w_k\left(S_k(\rho)-\bar S_{\rm post}(\rho)\right)^2}{\sum_k w_k}\right]^{1/2}.
\end{equation}
The corresponding prior width $\sigma_{\rm prior}(\rho)$ is obtained from the unweighted ensemble of the 30 parameter sets.
\begin{equation}
    \sigma_{\rm prior}(\rho)=\left[\frac{\sum_k \left(S_k(\rho)-\bar S_{\rm prior}(\rho)\right)^2}{30}\right]^{1/2}.
\end{equation}
}. 
The maximum value of $W(\rho)$, $W_{max}=0.498$, appears at 0.28$\rho_0$, where the corresponding symmetry energy within 1$\sigma_{\rm post}$ uncertainty is $S(0.28\rho_0)=13.84\pm 1.31$ MeV. When we take $W(\rho)/W_{max}>0.9$, the corresponding sensitive density region is $0.019\le\rho/\rho_0\le0.60$.

\begin{figure}[htbp]
	\centering
	\includegraphics[width=1.0\linewidth]{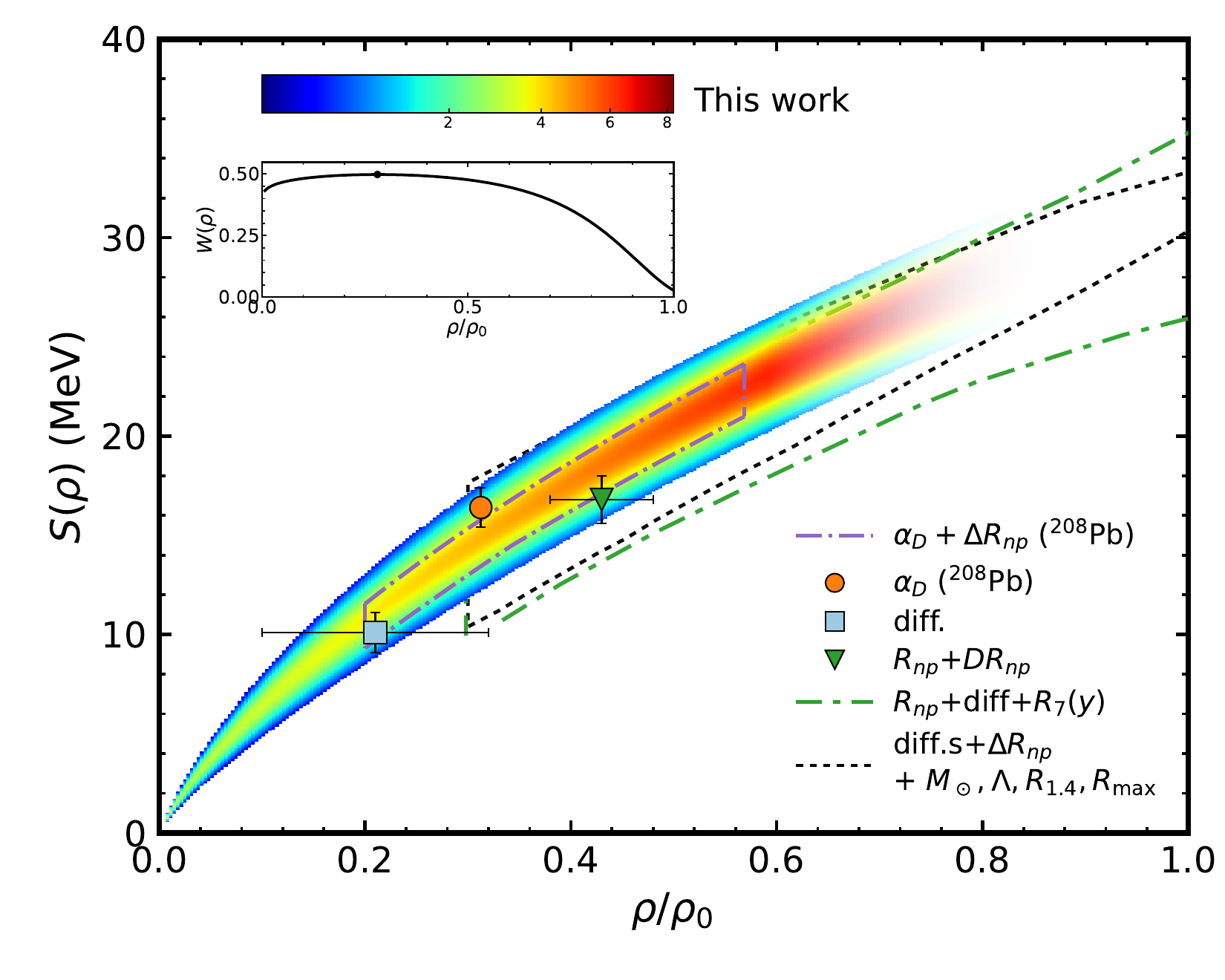}
	\caption{Relative posterior-distribution on the constraints of the density dependence of the symmetry energy. 
    The color denotes $P(y|\rho)$ 
    while the transparency controlled by $W(\rho)$ is used only as a visual guide to de-emphasize less informative density regions. The displayed band corresponds to $\bar S_{\rm post}(\rho)\pm2\sigma_{\rm post}(\rho)$.}
    \label{fig:constraint-symmetry-energy}
\end{figure}

Assuming the posterior-like distribution of $S$ at a given $\rho$ as
\begin{equation}
    P(S|\rho)=\frac{1}{\sqrt{2\pi}\sigma_{\rm post}(\rho)}\exp\left[-\frac{\left(S(\rho)-\bar S_{\rm post}(\rho)\right)^2}{2\sigma_{\rm post}^2(\rho)}\right],
\end{equation}
the constraint of symmetry energy can be visualized in the $S$--$\rho$ plane with a given confidence level. However, comparing $P(S|\rho)$ at different densities is not straightforward because the typical magnitude of $S$ varies significantly with $\rho$. For this purpose, it is useful to introduce the relative quantity $y(\rho)=S(\rho)/\bar{S}_{\rm post}(\rho)$ and the corresponding distribution $P(y|\rho)$ normalized as $\int P(y|\rho)dy=1$. Fig.~\ref{fig:constraint-symmetry-energy} shows $P(y|\rho)=\bar{S}_{\text{post}}(\rho)P(S|\rho)$ as a color map.  
The density region with weak narrowing strength, characterized by $W(\rho)<0.45$, is progressively faded through the opacity factor $\alpha(\rho)=[W(\rho)/0.45]^8$. This highlights that the present observables mainly constrain the symmetry energy below approximately $0.6\rho_0$, while providing only limited information at higher densities.

For comparison, several representative constraints from other isospin-sensitive observables are also displayed in Fig.~\ref{fig:constraint-symmetry-energy}. These include 
heavy-ion collision constraints
~\cite{Tsang09PRL,YXZhang20PRC,Lynch22PLB,Morfouace19PLB}, 
the simultaneous constraint from the RCNP dipole polarizability and the PREX-II neutron-skin result~\cite{DDNiu26arXiv}, and the Skyrme Hartree-Fock (SHF) + Random Phase Approximation (RPA) constraint from the electric dipole polarizability of $^{208}$Pb~\cite{JXu20PLB}. The present AMD constraint exhibits substantial overlap with these independent determinations in the density region around (0.2-0.6$\rho_0$), where the sensitivity of neutron-skin thicknesses and electric dipole polarizability is maximal. Such consistency is noteworthy because the present result is obtained from a unified time-dependent microscopic framework that simultaneously describes static density distributions and dynamical dipole responses. It therefore provides an independent validation of the symmetry-energy constraints inferred from EDF-based and heavy-ion approaches.

\section{Summary}\label{Sec:summary}

In summary, we have performed a systematic analysis of the neutron-skin thicknesses of nuclei from $^{40}\mathrm{Ca}$ to $^{238}\mathrm{U}$ together with the electric dipole polarizability of $^{208}\mathrm{Pb}$ within the AMD model. Thirty interaction parameter sets with different values of $S_0$, $L$, and the neutron-proton effective-mass splitting $\Delta m_{np}^*$ were used. 
We find that the neutron-skin thickness exhibits a clear mass-dependent enhancement of parameter sensitivity, with medium-heavy and heavy nuclei providing the strongest discrimination among the interactions. The dipole polarizability of $^{208}\mathrm{Pb}$ shows a pronounced sensitivity to the slope parameter $L$, while also retaining visible dependence on $S_0$ and effective-mass splitting. A combined $\chi^2$ analysis of neutron-skin thicknesses and electric dipole polarizability yields best-fit values of $L=53.0~{\rm MeV}$, $57.6~{\rm MeV}$, and $65.3~{\rm MeV}$ for $S_0=30$, 32, and 34~MeV, respectively, when $\Delta m_{np}^*<0$. For $\Delta m_{np}^*>0$, the corresponding values are $L=52.4~{\rm MeV}$, $55.4~{\rm MeV}$, and $62.1~{\rm MeV}$.

To quantitatively describe the sensitive density region probed by neutron-skin thicknesses and electric dipole polarizability, a relative narrowing strength function $W(\rho)$ is used. Our results show that the sensitive density region is $0.019\le \rho/\rho_0\le 0.60$, where the relative narrowing strength function varies by less than 10\% compared to its maximum narrowing strength. The maximum reduction of the uncertainty of $S(\rho)$ occurs at $\simeq0.28\rho_0$, where the symmetry energy within 1$\sigma_{\rm post}$ is constrained to be $S(0.28\rho_0)=13.84\pm 1.31$ MeV.
These results demonstrate that a unified AMD analysis of neutron-skin systematics and electric dipole polarizability provides a complementary constraint on the symmetry energy below saturation density.







\section*{Acknowledgements}

This work was partly supported by the National Natural Science Foundation of China under Grants No. 12275359, No. 12375129, No. 11875323 and No. 11961141003, by JSPS KAKENHI
Grant No. JP21K03528, by the National Key R\&D Program of China under Grant No. 2023 YFA1606402, by the Continuous Basic Scientific Research Project, by funding of the China Institute of Atomic Energy under Grant No. YZ222407001301, No. YZ232604001601, YC010270525794 and No. PA010271225779, and by the Leading Innovation Project of the CNNC under Grants No. LC192209000701 and No. LC202309000201. We acknowledge support by the computing server SCATP in China Institute of Atomic Energy.


\bibliographystyle{elsarticle-num}
\bibliography{ref}

@article{BALi08PR,
title = {Recent progress and new challenges in isospin physics with heavy-ion reactions},
journal = {Physics Reports},
volume = {464},
number = {4},
pages = {113-281},
year = {2008},
issn = {0370-1573},
doi = {https://doi.org/10.1016/j.physrep.2008.04.005},
url = {https://www.sciencedirect.com/science/article/pii/S0370157308001269},
author = {Bao-An Li and Lie-Wen Chen and Che Ming Ko}
}

@article{ROCA18PPNP,
title = {Nuclear equation of state from ground and collective excited state properties of nuclei},
journal = {Progress in Particle and Nuclear Physics},
volume = {101},
pages = {96-176},
year = {2018},
issn = {0146-6410},
doi = {https://doi.org/10.1016/j.ppnp.2018.04.001},
url = {https://www.sciencedirect.com/science/article/pii/S0146641018300334},
author = {X. Roca-Maza and N. Paar}
}

@article{Trippa08PRC,
  title = {Giant dipole resonance as a quantitative constraint on the symmetry energy},
  author = {Trippa, Luca and Col\`o, Gianluca and Vigezzi, Enrico},
  journal = {Phys. Rev. C},
  volume = {77},
  issue = {6},
  pages = {061304},
  numpages = {5},
  year = {2008},
  month = {Jun},
  publisher = {American Physical Society},
  doi = {10.1103/PhysRevC.77.061304},
  url = {https://link.aps.org/doi/10.1103/PhysRevC.77.061304}
}

@article{Roca13PRC,
  title = {Electric dipole polarizability in ${}^{208}$Pb: Insights from the droplet model},
  author = {Roca-Maza, X. and Brenna, M. and Col\`o, G. and Centelles, M. and Vi\~nas, X. and Agrawal, B. K. and Paar, N. and Vretenar, D. and Piekarewicz, J.},
  journal = {Phys. Rev. C},
  volume = {88},
  issue = {2},
  pages = {024316},
  numpages = {7},
  year = {2013},
  month = {Aug},
  publisher = {American Physical Society},
  doi = {10.1103/PhysRevC.88.024316},
  url = {https://link.aps.org/doi/10.1103/PhysRevC.88.024316}
}

@article{Tsang12PRC,
  title = {Constraints on the symmetry energy and neutron skins from experiments and theory},
  author = {Tsang, M. B. and Stone, J. R. and Camera, F. and Danielewicz, P. and Gandolfi, S. and Hebeler, K. and Horowitz, C. J. and Lee, Jenny and Lynch, W. G. and Kohley, Z. and Lemmon, R. and M\"oller, P. and Murakami, T. and Riordan, S. and Roca-Maza, X. and Sammarruca, F. and Steiner, A. W. and Vida\~na, I. and Yennello, S. J.},
  journal = {Phys. Rev. C},
  volume = {86},
  issue = {1},
  pages = {015803},
  numpages = {10},
  year = {2012},
  month = {Jul},
  publisher = {American Physical Society},
  doi = {10.1103/PhysRevC.86.015803},
  url = {https://link.aps.org/doi/10.1103/PhysRevC.86.015803}
}

@article{Reed21PRL,
  title = {Implications of PREX-2 on the Equation of State of Neutron-Rich Matter},
  author = {Reed, Brendan T. and Fattoyev, F. J. and Horowitz, C. J. and Piekarewicz, J.},
  journal = {Phys. Rev. Lett.},
  volume = {126},
  issue = {17},
  pages = {172503},
  numpages = {5},
  year = {2021},
  month = {Apr},
  publisher = {American Physical Society},
  doi = {10.1103/PhysRevLett.126.172503},
  url = {https://link.aps.org/doi/10.1103/PhysRevLett.126.172503}
}

@article{Reinhard10PRC,
  title = {Information content of a new observable: The case of the nuclear neutron skin},
  author = {Reinhard, P.-G. and Nazarewicz, W.},
  journal = {Phys. Rev. C},
  volume = {81},
  issue = {5},
  pages = {051303},
  numpages = {5},
  year = {2010},
  month = {May},
  publisher = {American Physical Society},
  doi = {10.1103/PhysRevC.81.051303},
  url = {https://link.aps.org/doi/10.1103/PhysRevC.81.051303}
}

@article{Brown2000PRL,
  title = {Neutron Radii in Nuclei and the Neutron Equation of State},
  author = {Alex Brown, B.},
  journal = {Phys. Rev. Lett.},
  volume = {85},
  issue = {25},
  pages = {5296--5299},
  numpages = {0},
  year = {2000},
  month = {Dec},
  publisher = {American Physical Society},
  doi = {10.1103/PhysRevLett.85.5296},
  url = {https://link.aps.org/doi/10.1103/PhysRevLett.85.5296}
}

@article{TGYue22PRR,
  title = {Constraints on the symmetry energy from PREX-II in the multimessenger era},
  author = {Yue, Tong-Gang and Chen, Lie-Wen and Zhang, Zhen and Zhou, Ying},
  journal = {Phys. Rev. Res.},
  volume = {4},
  issue = {2},
  pages = {L022054},
  numpages = {6},
  year = {2022},
  month = {Jun},
  publisher = {American Physical Society},
  doi = {10.1103/PhysRevResearch.4.L022054},
  url = {https://link.aps.org/doi/10.1103/PhysRevResearch.4.L022054}
}

@article{Lynch22PLB,
title = {Decoding the density dependence of the nuclear symmetry energy},
journal = {Physics Letters B},
volume = {830},
pages = {137098},
year = {2022},
issn = {0370-2693},
doi = {https://doi.org/10.1016/j.physletb.2022.137098},
url = {https://www.sciencedirect.com/science/article/pii/S0370269322002325},
author = {W.G. Lynch and M.B. Tsang}
}

@article{Morfouace19PLB,
  title = {Constraining the symmetry energy with heavy-ion collisions and Bayesian analyses},
  volume = {799},
  ISSN = {0370-2693},
  url = {http://dx.doi.org/10.1016/j.physletb.2019.135045},
  DOI = {10.1016/j.physletb.2019.135045},
  journal = {Physics Letters B},
  publisher = {Elsevier BV},
  author = {Morfouace,  P. and Tsang,  C.Y. and Zhang,  Y. and Lynch,  W.G. and Tsang,  M.B. and Coupland,  D.D.S. and Youngs,  M. and Chajecki,  Z. and Famiano,  M.A. and Ghosh,  T.K. and Jhang,  G. and Lee,  Jenny and Liu,  H. and Sanetullaev,  A. and Showalter,  R. and Winkelbauer,  J.},
  year = {2019},
  month = dec,
  pages = {135045}
}

@article{JXu20PLB,
title = {Constraining isovector nuclear interactions with giant resonances within a Bayesian approach},
journal = {Physics Letters B},
volume = {810},
pages = {135820},
year = {2020},
issn = {0370-2693},
doi = {https://doi.org/10.1016/j.physletb.2020.135820},
url = {https://www.sciencedirect.com/science/article/pii/S0370269320306237},
author = {Jun Xu and Jia Zhou and Zhen Zhang and Wen-Jie Xie and Bao-An Li}
}

@article{Centelles09PRL,
  title = {Nuclear Symmetry Energy Probed by Neutron Skin Thickness of Nuclei},
  author = {Centelles, M. and Roca-Maza, X. and Vi\~nas, X. and Warda, M.},
  journal = {Phys. Rev. Lett.},
  volume = {102},
  issue = {12},
  pages = {122502},
  numpages = {4},
  year = {2009},
  month = {Mar},
  publisher = {American Physical Society},
  doi = {10.1103/PhysRevLett.102.122502},
  url = {https://link.aps.org/doi/10.1103/PhysRevLett.102.122502}
}

@article{Tsang24nature,
  title = {Determination of the equation of state from nuclear experiments and neutron star observations},
  volume = {8},
  ISSN = {2397-3366},
  url = {http://dx.doi.org/10.1038/s41550-023-02161-z},
  DOI = {10.1038/s41550-023-02161-z},
  number = {3},
  journal = {Nature Astronomy},
  publisher = {Springer Science and Business Media LLC},
  author = {Tsang,  Chun Yuen and Tsang,  ManYee Betty and Lynch,  William G. and Kumar,  Rohit and Horowitz,  Charles J.},
  year = {2024},
  month = jan,
  pages = {328–336}
}

@article{Lattimer14EPJ,
  title = {Constraints on the symmetry energy using the mass-radius relation of neutron stars},
  volume = {50},
  ISSN = {1434-601X},
  url = {http://dx.doi.org/10.1140/epja/i2014-14040-y},
  DOI = {10.1140/epja/i2014-14040-y},
  number = {2},
  journal = {The European Physical Journal A},
  publisher = {Springer Science and Business Media LLC},
  author = {Lattimer,  James M. and Steiner,  Andrew W.},
  year = {2014},
  month = feb 
}

@article{NBZhang18APJ,
doi = {10.3847/1538-4357/aac027},
url = {https://doi.org/10.3847/1538-4357/aac027},
year = {2018},
month = {may},
publisher = {The American Astronomical Society},
volume = {859},
number = {2},
pages = {90},
author = {Zhang, Nai-Bo and Li, Bao-An and Xu, Jun},
title = {Combined Constraints on the Equation of State of Dense Neutron-rich Matter from Terrestrial Nuclear Experiments and Observations of Neutron Stars},
journal = {The Astrophysical Journal}
}

@article{Tsang09PRL,
  title = {Constraints on the Density Dependence of the Symmetry Energy},
  author = {Tsang, M. B. and Zhang, Yingxun and Danielewicz, P. and Famiano, M. and Li, Zhuxia and Lynch, W. G. and Steiner, A. W.},
  journal = {Phys. Rev. Lett.},
  volume = {102},
  issue = {12},
  pages = {122701},
  numpages = {4},
  year = {2009},
  month = {Mar},
  publisher = {American Physical Society},
  doi = {10.1103/PhysRevLett.102.122701},
  url = {https://link.aps.org/doi/10.1103/PhysRevLett.102.122701}
}

@article{LWChen05PRL,
  title = {Determination of the Stiffness of the Nuclear Symmetry Energy from Isospin Diffusion},
  author = {Chen, Lie-Wen and Ko, Che Ming and Li, Bao-An},
  journal = {Phys. Rev. Lett.},
  volume = {94},
  issue = {3},
  pages = {032701},
  numpages = {4},
  year = {2005},
  month = {Jan},
  publisher = {American Physical Society},
  doi = {10.1103/PhysRevLett.94.032701},
  url = {https://link.aps.org/doi/10.1103/PhysRevLett.94.032701}
}

@article{LGCao15PRC,
  title = {Constraints on the neutron skin and symmetry energy from the anti-analog giant dipole resonance in $^{208}\mathrm{Pb}$},
  author = {Cao, Li-Gang and Roca-Maza, X. and Col\`o, G. and Sagawa, H.},
  journal = {Phys. Rev. C},
  volume = {92},
  issue = {3},
  pages = {034308},
  numpages = {10},
  year = {2015},
  month = {Sep},
  publisher = {American Physical Society},
  doi = {10.1103/PhysRevC.92.034308},
  url = {https://link.aps.org/doi/10.1103/PhysRevC.92.034308}
}

@article{Roca11PRL,
  title = {Neutron Skin of $^{208}\mathrm{Pb}$, Nuclear Symmetry Energy, and the Parity Radius Experiment},
  author = {Roca-Maza, X. and Centelles, M. and Vi\~nas, X. and Warda, M.},
  journal = {Phys. Rev. Lett.},
  volume = {106},
  issue = {25},
  pages = {252501},
  numpages = {4},
  year = {2011},
  month = {Jun},
  publisher = {American Physical Society},
  doi = {10.1103/PhysRevLett.106.252501},
  url = {https://link.aps.org/doi/10.1103/PhysRevLett.106.252501}
}

@article{ZZhang13PLB,
  title = {Constraining the symmetry energy at subsaturation densities using isotope binding energy difference and neutron skin thickness},
  volume = {726},
  ISSN = {0370-2693},
  url = {http://dx.doi.org/10.1016/j.physletb.2013.08.002},
  DOI = {10.1016/j.physletb.2013.08.002},
  number = {1–3},
  journal = {Physics Letters B},
  publisher = {Elsevier BV},
  author = {Zhang,  Zhen and Chen,  Lie-Wen},
  year = {2013},
  month = oct,
  pages = {234–238}
}

@article{WBHe14PRL,
  title = {Giant Dipole Resonance as a Fingerprint of $\ensuremath{\alpha}$ Clustering Configurations in $^{12}\mathrm{C}$ and $^{16}\mathrm{O}$},
  author = {He, W. B. and Ma, Y. G. and Cao, X. G. and Cai, X. Z. and Zhang, G. Q.},
  journal = {Phys. Rev. Lett.},
  volume = {113},
  issue = {3},
  pages = {032506},
  numpages = {6},
  year = {2014},
  month = {Jul},
  publisher = {American Physical Society},
  doi = {10.1103/PhysRevLett.113.032506},
  url = {https://link.aps.org/doi/10.1103/PhysRevLett.113.032506}
}

@article{Reinhard21PRL,
  title = {Information Content of the Parity-Violating Asymmetry in $^{208}\mathrm{Pb}$},
  author = {Reinhard, Paul-Gerhard and Roca-Maza, Xavier and Nazarewicz, Witold},
  journal = {Phys. Rev. Lett.},
  volume = {127},
  issue = {23},
  pages = {232501},
  numpages = {7},
  year = {2021},
  month = {Nov},
  publisher = {American Physical Society},
  doi = {10.1103/PhysRevLett.127.232501},
  url = {https://link.aps.org/doi/10.1103/PhysRevLett.127.232501}
}

@article{Natsumi16PRC,
  title = {Probing neutron-proton dynamics by pions},
  author = {Ikeno, Natsumi and Ono, Akira and Nara, Yasushi and Ohnishi, Akira},
  journal = {Phys. Rev. C},
  volume = {93},
  issue = {4},
  pages = {044612},
  numpages = {13},
  year = {2016},
  month = {Apr},
  publisher = {American Physical Society},
  doi = {10.1103/PhysRevC.93.044612},
  url = {https://link.aps.org/doi/10.1103/PhysRevC.93.044612}
}

@article{Adhikari21PRL,
  title = {Accurate Determination of the Neutron Skin Thickness of $^{208}\mathrm{Pb}$ through Parity-Violation in Electron Scattering},
  author = {Adhikari, D. and Albataineh, H. and Androic, D. and Aniol, K. and Armstrong, D. S. and Averett, T. and Ayerbe Gayoso, C. and Barcus, S. and Bellini, V. and Beminiwattha, R. S. and Benesch, J. F. and Bhatt, H. and Bhatta Pathak, D. and Bhetuwal, D. and Blaikie, B. and Campagna, Q. and Camsonne, A. and Cates, G. D. and Chen, Y. and Clarke, C. and Cornejo, J. C. and Covrig Dusa, S. and Datta, P. and Deshpande, A. and Dutta, D. and Feldman, C. and Fuchey, E. and Gal, C. and Gaskell, D. and Gautam, T. and Gericke, M. and Ghosh, C. and Halilovic, I. and Hansen, J.-O. and Hauenstein, F. and Henry, W. and Horowitz, C. J. and Jantzi, C. and Jian, S. and Johnston, S. and Jones, D. C. and Karki, B. and Katugampola, S. and Keppel, C. and King, P. M. and King, D. E. and Knauss, M. and Kumar, K. S. and Kutz, T. and Lashley-Colthirst, N. and Leverick, G. and Liu, H. and Liyange, N. and Malace, S. and Mammei, R. and Mammei, J. and McCaughan, M. and McNulty, D. and Meekins, D. and Metts, C. and Michaels, R. and Mondal, M. M. and Napolitano, J. and Narayan, A. and Nikolaev, D. and Rashad, M. N. H. and Owen, V. and Palatchi, C. and Pan, J. and Pandey, B. and Park, S. and Paschke, K. D. and Petrusky, M. and Pitt, M. L. and Premathilake, S. and Puckett, A. J. R. and Quinn, B. and Radloff, R. and Rahman, S. and Rathnayake, A. and Reed, B. T. and Reimer, P. E. and Richards, R. and Riordan, S. and Roblin, Y. and Seeds, S. and Shahinyan, A. and Souder, P. and Tang, L. and Thiel, M. and Tian, Y. and Urciuoli, G. M. and Wertz, E. W. and Wojtsekhowski, B. and Yale, B. and Ye, T. and Yoon, A. and Zec, A. and Zhang, W. and Zhang, J. and Zheng, X.},
  collaboration = {PREX Collaboration},
  journal = {Phys. Rev. Lett.},
  volume = {126},
  issue = {17},
  pages = {172502},
  numpages = {7},
  year = {2021},
  month = {Apr},
  publisher = {American Physical Society},
  doi = {10.1103/PhysRevLett.126.172502},
  url = {https://link.aps.org/doi/10.1103/PhysRevLett.126.172502}
}

@article{YXZhang14PLB,
title = {Constraints on nucleon effective mass splitting with heavy ion collisions},
journal = {Physics Letters B},
volume = {732},
pages = {186-190},
year = {2014},
issn = {0370-2693},
doi = {https://doi.org/10.1016/j.physletb.2014.03.030},
url = {https://www.sciencedirect.com/science/article/pii/S0370269314001865},
author = {Yingxun Zhang and M.B. Tsang and Zhuxia Li and Hang Liu}
}

@article{Lattimer12AR,
   author = "Lattimer, James M.",
   title = "The Nuclear Equation of State and Neutron Star Masses", 
   journal= "Annual Review of Nuclear and Particle Science",
   year = "2012",
   volume = "62",
   number = "Volume 62, 2012",
   pages = "485-515",
   doi = "https://doi.org/10.1146/annurev-nucl-102711-095018",
   url = "https://www.annualreviews.org/content/journals/10.1146/annurev-nucl-102711-095018",
   publisher = "Annual Reviews",
   issn = "1545-4134",
   type = "Journal Article",
  }

@article{LWChen10PRC,
  title = {Density slope of the nuclear symmetry energy from the neutron skin thickness of heavy nuclei},
  author = {Chen, Lie-Wen and Ko, Che Ming and Li, Bao-An and Xu, Jun},
  journal = {Phys. Rev. C},
  volume = {82},
  issue = {2},
  pages = {024321},
  numpages = {7},
  year = {2010},
  month = {Aug},
  publisher = {American Physical Society},
  doi = {10.1103/PhysRevC.82.024321},
  url = {https://link.aps.org/doi/10.1103/PhysRevC.82.024321}
}

@article{YXZhang20PRC,
  title = {Constraints on the symmetry energy and its associated parameters from nuclei to neutron stars},
  author = {Zhang, Yingxun and Liu, Min and Xia, Cheng-Jun and Li, Zhuxia and Biswal, S. K.},
  journal = {Phys. Rev. C},
  volume = {101},
  issue = {3},
  pages = {034303},
  numpages = {11},
  year = {2020},
  month = {Mar},
  publisher = {American Physical Society},
  doi = {10.1103/PhysRevC.101.034303},
  url = {https://link.aps.org/doi/10.1103/PhysRevC.101.034303}
}

@article{Burgio21PPNP,
title = {Neutron stars and the nuclear equation of state},
journal = {Progress in Particle and Nuclear Physics},
volume = {120},
pages = {103879},
year = {2021},
issn = {0146-6410},
doi = {https://doi.org/10.1016/j.ppnp.2021.103879},
url = {https://www.sciencedirect.com/science/article/pii/S0146641021000338},
author = {G.F. Burgio and H.-J. Schulze and I. Vidaña and J.-B. Wei}
}

@misc{NUPECC24LRP,
  title = "{NuPECC Long Range Plan 2024 for European Nuclear Physics}",
  collaboration = "{Nuclear Physics European Committee}",
  year = 2025,
  eprint = {2503.15575},
  archivePrefix = {arXiv},
  primaryClass = {nucl-ex},
}

@techreport{USDOE23LRP,
  author = {{US Department of Energy (USDOE)}},
  title = {A New Era of Discovery: The 2023 Long Range Plan for Nuclear Science},
  institution = {US Department of Energy (USDOE)},
  year = {2023},
  doi = {10.2172/2280968},
  url = {https://www.osti.gov/biblio/2280968}
}

@article{BALi21Universe,
  title = {Progress in Constraining Nuclear Symmetry Energy Using Neutron Star Observables Since GW170817},
  volume = {7},
  ISSN = {2218-1997},
  url = {http://dx.doi.org/10.3390/universe7060182},
  DOI = {10.3390/universe7060182},
  number = {6},
  journal = {Universe},
  publisher = {MDPI AG},
  author = {Li,  Bao-An and Cai,  Bao-Jun and Xie,  Wen-Jie and Zhang,  Nai-Bo},
  year = {2021},
  month = jun,
  pages = {182}
}

@article{Ono92PRL,
  title = {Fragment formation studied with antisymmetrized version of molecular dynamics with two-nucleon collisions},
  author = {Ono, A. and Horiuchi, H. and Maruyama, T. and Ohnishi, A.},
  journal = {Phys. Rev. Lett.},
  volume = {68},
  issue = {19},
  pages = {2898--2900},
  numpages = {0},
  year = {1992},
  month = {May},
  publisher = {American Physical Society},
  doi = {10.1103/PhysRevLett.68.2898},
  url = {https://link.aps.org/doi/10.1103/PhysRevLett.68.2898}
}

@article{Ono92PTP,
    author = {Ono, Akira and Horiuchi, Hisashi and Maruyama, Toshiki and Ohnishi, Akira},
    title = {Antisymmetrized Version of Molecular Dynamics with Two-Nucleon Collisions and Its Application to Heavy Ion Reactions},
    journal = {Progress of Theoretical Physics},
    volume = {87},
    number = {5},
    pages = {1185-1206},
    year = {1992},
    month = {05},
    issn = {0033-068X},
    doi = {10.1143/ptp/87.5.1185},
    url = {https://doi.org/10.1143/ptp/87.5.1185},
    eprint = {https://academic.oup.com/ptp/article-pdf/87/5/1185/5272175/87-5-1185.pdf},
}

@article{Vretenar12PRC,
  title = {Low-energy isovector and isoscalar dipole response in neutron-rich nuclei},
  author = {Vretenar, D. and Niu, Y. F. and Paar, N. and Meng, J.},
  journal = {Phys. Rev. C},
  volume = {85},
  issue = {4},
  pages = {044317},
  numpages = {8},
  year = {2012},
  month = {Apr},
  publisher = {American Physical Society},
  doi = {10.1103/PhysRevC.85.044317},
  url = {https://link.aps.org/doi/10.1103/PhysRevC.85.044317}
}

@article{Tamii11PRL,
  title = {Complete Electric Dipole Response and the Neutron Skin in $^{208}\mathrm{Pb}$},
  author = {Tamii, A. and Poltoratska, I. and von Neumann-Cosel, P. and Fujita, Y. and Adachi, T. and Bertulani, C. A. and Carter, J. and Dozono, M. and Fujita, H. and Fujita, K. and Hatanaka, K. and Ishikawa, D. and Itoh, M. and Kawabata, T. and Kalmykov, Y. and Krumbholz, A. M. and Litvinova, E. and Matsubara, H. and Nakanishi, K. and Neveling, R. and Okamura, H. and Ong, H. J. and \"Ozel-Tashenov, B. and Ponomarev, V. Yu. and Richter, A. and Rubio, B. and Sakaguchi, H. and Sakemi, Y. and Sasamoto, Y. and Shimbara, Y. and Shimizu, Y. and Smit, F. D. and Suzuki, T. and Tameshige, Y. and Wambach, J. and Yamada, R. and Yosoi, M. and Zenihiro, J.},
  journal = {Phys. Rev. Lett.},
  volume = {107},
  issue = {6},
  pages = {062502},
  numpages = {5},
  year = {2011},
  month = {Aug},
  publisher = {American Physical Society},
  doi = {10.1103/PhysRevLett.107.062502},
  url = {https://link.aps.org/doi/10.1103/PhysRevLett.107.062502}
}

@article{Klimkiewicz07PRC,
  title = {Nuclear symmetry energy and neutron skins derived from pygmy dipole resonances},
  author = {Klimkiewicz, A. and Paar, N. and Adrich, P. and Fallot, M. and Boretzky, K. and Aumann, T. and Cortina-Gil, D. and Pramanik, U. Datta and Elze, Th. W. and Emling, H. and Geissel, H. and Hellstr\"om, M. and Jones, K. L. and Kratz, J. V. and Kulessa, R. and Nociforo, C. and Palit, R. and Simon, H. and Sur\'owka, G. and S\"ummerer, K. and Vretenar, D. and Walu\ifmmode \acute{s}\else \'{s}\fi{}, W.},
  collaboration = {LAND Collaboration},
  journal = {Phys. Rev. C},
  volume = {76},
  issue = {5},
  pages = {051603},
  numpages = {4},
  year = {2007},
  month = {Nov},
  publisher = {American Physical Society},
  doi = {10.1103/PhysRevC.76.051603},
  url = {https://link.aps.org/doi/10.1103/PhysRevC.76.051603}
}

@article{Carbone10PRC,
  title = {Constraints on the symmetry energy and neutron skins from pygmy resonances in $^{68}\mathrm{Ni}$ and $^{132}\mathrm{Sn}$},
  author = {Carbone, Andrea and Col\`o, Gianluca and Bracco, Angela and Cao, Li-Gang and Bortignon, Pier Francesco and Camera, Franco and Wieland, Oliver},
  journal = {Phys. Rev. C},
  volume = {81},
  issue = {4},
  pages = {041301},
  numpages = {5},
  year = {2010},
  month = {Apr},
  publisher = {American Physical Society},
  doi = {10.1103/PhysRevC.81.041301},
  url = {https://link.aps.org/doi/10.1103/PhysRevC.81.041301}
}

@article{Colo14EPJ,
  title = {Symmetry energy from the nuclear collective motion: constraints from dipole,  quadrupole,  monopole and spin-dipole resonances},
  volume = {50},
  ISSN = {1434-601X},
  url = {http://dx.doi.org/10.1140/epja/i2014-14026-9},
  DOI = {10.1140/epja/i2014-14026-9},
  number = {2},
  journal = {The European Physical Journal A},
  publisher = {Springer Science and Business Media LLC},
  author = {Colò,  G. and Garg,  U. and Sagawa,  H.},
  year = {2014},
  month = feb 
}

@article{HZheng16PRC,
  title = {Dipole response in neutron-rich nuclei with new Skyrme interactions},
  author = {Zheng, H. and Burrello, S. and Colonna, M. and Baran, V.},
  journal = {Phys. Rev. C},
  volume = {94},
  issue = {1},
  pages = {014313},
  numpages = {14},
  year = {2016},
  month = {Jul},
  publisher = {American Physical Society},
  doi = {10.1103/PhysRevC.94.014313},
  url = {https://link.aps.org/doi/10.1103/PhysRevC.94.014313}
}

@article{Natsumi23PRC,
  title = {Collision integral with momentum-dependent potentials and its impact on pion production in heavy-ion collisions},
  author = {Ikeno, Natsumi and Ono, Akira},
  journal = {Phys. Rev. C},
  volume = {108},
  issue = {4},
  pages = {044601},
  numpages = {18},
  year = {2023},
  month = {Oct},
  publisher = {American Physical Society},
  doi = {10.1103/PhysRevC.108.044601},
  url = {https://link.aps.org/doi/10.1103/PhysRevC.108.044601}
}

@article{Adhikari22PRL,
  title = {Precision Determination of the Neutral Weak Form Factor of $^{48}\mathrm{Ca}$},
  author = {Adhikari, D. and Albataineh, H. and Androic, D. and Aniol, K. A. and Armstrong, D. S. and Averett, T. and Ayerbe Gayoso, C. and Barcus, S. K. and Bellini, V. and Beminiwattha, R. S. and Benesch, J. F. and Bhatt, H. and Bhatta Pathak, D. and Bhetuwal, D. and Blaikie, B. and Boyd, J. and Campagna, Q. and Camsonne, A. and Cates, G. D. and Chen, Y. and Clarke, C. and Cornejo, J. C. and Covrig Dusa, S. and Dalton, M. M. and Datta, P. and Deshpande, A. and Dutta, D. and Feldman, C. and Fuchey, E. and Gal, C. and Gaskell, D. and Gautam, T. and Gericke, M. and Ghosh, C. and Halilovic, I. and Hansen, J.-O. and Hassan, O. and Hauenstein, F. and Henry, W. and Horowitz, C. J. and Jantzi, C. and Jian, S. and Johnston, S. and Jones, D. C. and Kakkar, S. and Katugampola, S. and Keppel, C. and King, P. M. and King, D. E. and Kumar, K. S. and Kutz, T. and Lashley-Colthirst, N. and Leverick, G. and Liu, H. and Liyanage, N. and Mammei, J. and Mammei, R. and McCaughan, M. and McNulty, D. and Meekins, D. and Metts, C. and Michaels, R. and Mihovilovic, M. and Mondal, M. M. and Napolitano, J. and Narayan, A. and Nikolaev, D. and Owen, V. and Palatchi, C. and Pan, J. and Pandey, B. and Park, S. and Paschke, K. D. and Petrusky, M. and Pitt, M. L. and Premathilake, S. and Quinn, B. and Radloff, R. and Rahman, S. and Rashad, M. N. H. and Rathnayake, A. and Reed, B. T. and Reimer, P. E. and Richards, R. and Riordan, S. and Roblin, Y. R. and Seeds, S. and Shahinyan, A. and Souder, P. and Thiel, M. and Tian, Y. and Urciuoli, G. M. and Wertz, E. W. and Wojtsekhowski, B. and Yale, B. and Ye, T. and Yoon, A. and Xiong, W. and Zec, A. and Zhang, W. and Zhang, J. and Zheng, X.},
  collaboration = {CREX Collaboration},
  journal = {Phys. Rev. Lett.},
  volume = {129},
  issue = {4},
  pages = {042501},
  numpages = {8},
  year = {2022},
  month = {Jul},
  publisher = {American Physical Society},
  doi = {10.1103/PhysRevLett.129.042501},
  url = {https://link.aps.org/doi/10.1103/PhysRevLett.129.042501}
}

@article{TGYue26Science,
   title={Evidence for strong isovector nuclear spin–orbit interaction},
   volume={71},
   ISSN={2095-9273},
   url={http://dx.doi.org/10.1016/j.scib.2026.01.062},
   DOI={10.1016/j.scib.2026.01.062},
   number={6},
   journal={Science Bulletin},
   publisher={Elsevier BV},
   author={Yue, Tong-Gang and Zhang, Zhen and Chen, Lie-Wen},
   year={2026},
   month=mar, pages={1270–1272} }

@misc{DDNiu26arXiv,
      title={Constraining the nuclear symmetry energy from electric dipole polarizability and neutron skin in $^{208}\mathrm{Pb}$ within antisymmetrized molecular dynamics}, 
      author={Dandan Niu and Xinyu Wang and Ying Cui and Qiang Zhao and Kai Zhao and Akira Ono and Yingxun Zhang},
      year={2026},
      eprint={2602.19039},
      archivePrefix={arXiv},
      primaryClass={nucl-th},
      url={https://arxiv.org/abs/2602.19039}, 
}

@article{Bennaceur05CPC,
title = {Coordinate-space solution of the Skyrme–Hartree–Fock– Bogolyubov equations within spherical symmetry. The program HFBRAD (v1.00)},
journal = {Computer Physics Communications},
volume = {168},
number = {2},
pages = {96-122},
year = {2005},
issn = {0010-4655},
doi = {https://doi.org/10.1016/j.cpc.2005.02.002},
url = {https://www.sciencedirect.com/science/article/pii/S0010465505002304},
author = {K. Bennaceur and J. Dobaczewski}
}

@article{Pawel17NPA,
title = {Symmetry energy III: Isovector skins},
journal = {Nuclear Physics A},
volume = {958},
pages = {147-186},
year = {2017},
issn = {0375-9474},
doi = {https://doi.org/10.1016/j.nuclphysa.2016.11.008},
url = {https://www.sciencedirect.com/science/article/pii/S0375947416302895},
author = {Paweł Danielewicz and Pardeep Singh and Jenny Lee}
}

@article{Kortelainen10PRC,
  title = {Nuclear energy density optimization},
  author = {Kortelainen, M. and Lesinski, T. and Mor\'e, J. and Nazarewicz, W. and Sarich, J. and Schunck, N. and Stoitsov, M. V. and Wild, S.},
  journal = {Phys. Rev. C},
  volume = {82},
  issue = {2},
  pages = {024313},
  numpages = {18},
  year = {2010},
  month = {Aug},
  publisher = {American Physical Society},
  doi = {10.1103/PhysRevC.82.024313},
  url = {https://link.aps.org/doi/10.1103/PhysRevC.82.024313}
}

@article{kunji,
  title = {Role of the isovector spin-orbit potential in mitigating the CREX-PREX dilemma},
  author = {Kunjipurayil, Athul and Piekarewicz, J. and Salinas, Marc},
  journal = {Phys. Rev. C},
  volume = {112},
  issue = {1},
  pages = {014310},
  numpages = {10},
  year = {2025},
  month = {Jul},
  publisher = {American Physical Society},
  doi = {10.1103/tcy2-brmk},
  url = {https://link.aps.org/doi/10.1103/tcy2-brmk}
}

@misc{Piekarewicz26MREX,
      title={The Matter Radius of 132Sn and the CREX-PREX Dilemma}, 
      author={J. Piekarewicz},
      year={2026},
      eprint={2603.11983},
      archivePrefix={arXiv},
      primaryClass={nucl-th},
      url={https://arxiv.org/abs/2603.11983}, 
}

@article{Trzciska01PRL,
  title = {Neutron Density Distributions Deduced from Antiprotonic Atoms},
  volume = {87},
  ISSN = {1079-7114},
  url = {http://dx.doi.org/10.1103/PhysRevLett.87.082501},
  DOI = {10.1103/physrevlett.87.082501},
  number = {8},
  journal = {Physical Review Letters},
  publisher = {American Physical Society (APS)},
  author = {Trzci{\'n}ska,  A. and Jastrz{\k{e}}bski,  J. and Lubi{\'n}ski,  P. and Hartmann,  F. J. and Schmidt,  R. and von Egidy,  T. and k{\l}os,  B.},
  year = {2001},
  month = Aug 
}


\end{document}